\documentclass[conference]{IEEEtran}
\IEEEoverridecommandlockouts
\usepackage{graphicx}
\usepackage{cite}
\usepackage{float}
\usepackage{multirow}
\usepackage[ruled,vlined,linesnumbered]{algorithm2e}
\usepackage{ragged2e}
\usepackage{amsmath}
\usepackage[table,xcdraw]{xcolor}
\usepackage{booktabs}
\usepackage{hyperref}
\hypersetup{
    colorlinks = true, 
    urlcolor = black, 
    linkcolor = blue, 
    citecolor = blue 
}

\renewcommand{\thetable}{\arabic{table}}

\def\tablename{Table}
\def\figurename{Figure}

\begin{document}

\title{Exploring the Potential of\\Wireless-enabled Multi-Chip AI Accelerators}

\author{
    \IEEEauthorblockN{
        Emmanuel Irabor,
        Mariam Musavi,
        Abhijit Das and
        Sergi Abadal
    }
    \IEEEauthorblockA{
        \textit{Universitat Polit\`ecnica de Catalunya, Barcelona, Spain}\\
        emmanuel.onyekachukwu.irabor@estudiantat.upc.edu
    }
    \thanks{Authors gratefully acknowledge funding from the European Commission through projects with GA 101042080 (WINC) and 101189474 (EWiC).}
}

\maketitle

\begin{abstract}
The insatiable appetite of Artificial Intelligence (AI) workloads for computing power is pushing the industry to develop faster and more efficient accelerators. The rigidity of custom hardware, however, conflicts with the need for scalable and versatile architectures capable of catering to the needs of the evolving and heterogeneous pool of Machine Learning (ML) models in the literature. In this context, multi-chiplet architectures assembling multiple (perhaps heterogeneous) accelerators are an appealing option that is unfortunately hindered by the still rigid and inefficient chip-to-chip interconnects. In this paper, we explore the potential of wireless technology as a complement to existing wired interconnects in this multi-chiplet approach. Using an evaluation framework from the state-of-the-art, we show that wireless interconnects can lead to speedups of 10\% on average and 20\% maximum. We also highlight the importance of load balancing between the wired and wireless interconnects, which will be further explored in future work.
\end{abstract}


\section{Introduction}
AI has revolutionized a wide range of fields thanks to its superhuman classification, discrimination, and generative capabilities~\cite{lecun2015deep,lin2022survey,hoffmann2022training}. However, the continuous advancements made in the different ML models that sustain such a revolution also lead to a constant increase in their computational requirements. Indeed, due to their evolving size and diversity, modern ML models urgently call for faster, more efficient, and more flexible computing platforms~\cite{sze2020efficient}. 

To address the speed and efficiency issues, a wide range of specialized hardware accelerators have been presented in the last decade~\cite{chen2019eyeriss, kwon2021heterogeneous, abadal2021computing}. These accelerators are typically composed of a large array of Processing Elements (PEs, generally in the form of multiply-accumulate units) implementing a given fixed dataflow through a dense Network-on-Chip (NoC)~\cite{chatarasi2021marvel}. These architectures go beyond the performance and energy efficiency of Graphical Processing Units (GPUs), yet at the cost of a loss of generality or versatility. Indeed, it is extremely challenging to scale and reconfigure such AI accelerators to execute ever-growing and heterogeneous AI workloads without sacrificing performance and efficiency~\cite{kwon2018maeri,yang2023versa}. 

\begin{figure}[t]
\includegraphics[width=\columnwidth]{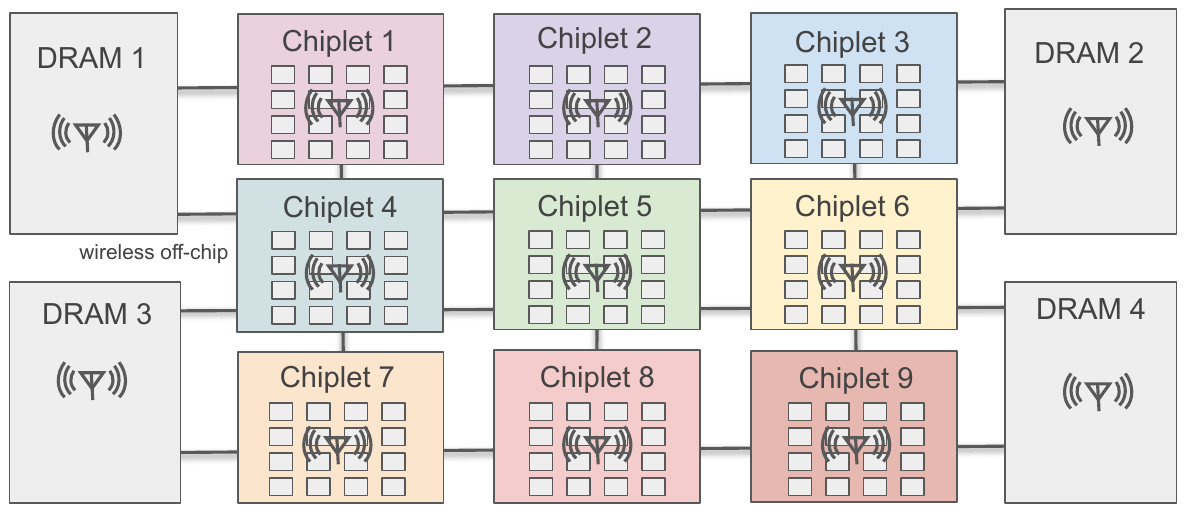}
\caption{A schematic architecture of wireless-enabled multi-chip AI accelerator with 3$\times$3 chiplets and 4 DRAMs. An antenna and transceiver are integrated at the center of each DRAM and compute chiplet.}
\label{fig:schematic}
\end{figure}

Chiplet technology is a promising approach that could enable the creation of scalable and versatile AI accelerators, by combining together multiple specialized (and potentially heterogeneous) AI accelerator chiplets in a single computing platform, as illustrated in Figure~\ref{fig:schematic}. These chiplets are interconnected among themselves and to memory via on-package links, typically through silicon interposers or organic substrates, in order to create a Network-on-Package (NoP) \cite{beck2018zeppelin, kannan2015enabling, vivet20202}. Hence, chiplet-based architectures show a promising path to address the scaling challenge of computing platforms, as hinted in multiple works, including SIMBA \cite{shao2019simba} or WIENNA \cite{guirado2021dataflow}.

One of the main drawbacks of multi-chiplet architectures is the reduction of the interconnect performance and efficiency, which is crucial when serving workloads that imply significant data movement across the architecture \cite{boroumand2018google, das2024chip}.
This is not only due to the limited speed of memory modules but, more critically, the relatively slow chiplet-to-chiplet data transfers. These data transfers often dominate the system energy as they rely on traversing the long interconnects. The problem is further intensified by the use of collective (i.e. multicast and reduction) communication in many dataflows employed by AI accelerators \cite{musavi2024, guirado2021dataflow, sze2020efficient}.

\begin{figure}[t]
\includegraphics[width=\columnwidth]{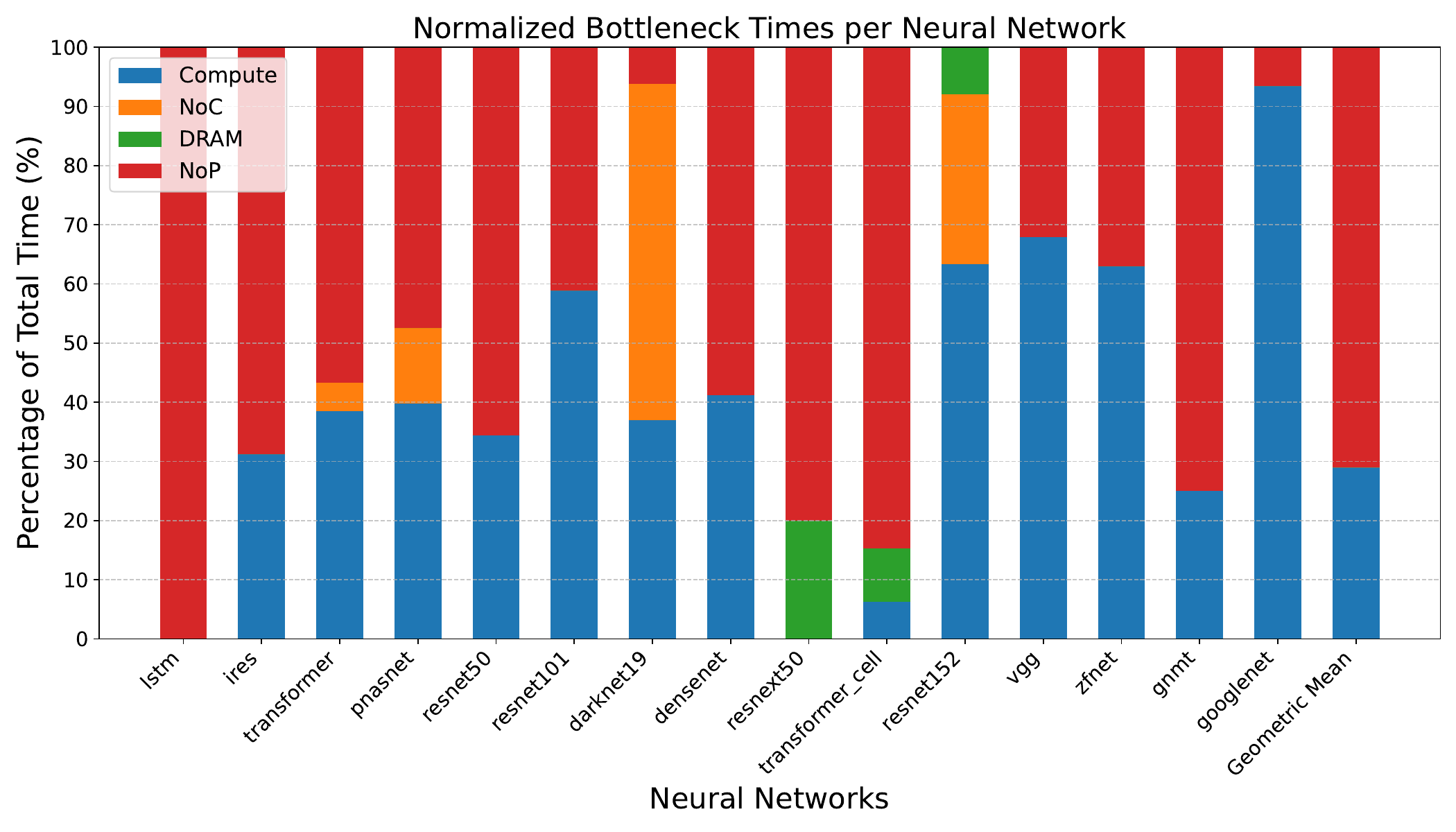}
\caption{Percentage of time where each of the elements of a 144-TOPS 3$\times$3 multi-chip AI accelerator is the performance bottleneck.}
\label{fig:bottleneck}
\end{figure}

To illustrate the point made above, we simulated multiple AI workloads over a 144-TOPS accelerator broken down in 3$\times$3 chiplets using the methods described later in Section~\ref{sec:methodology} and summarized in Table~\ref{tab:simulationParameters-label}. We recorded, for each of the layers of the workload, its execution time and which is the performance bottleneck. Figure~\ref{fig:bottleneck} summarizes the results, showing the percentage of the time that each element of the architecture is a bottleneck. The results clearly indicate that the chiplet interconnects (i.e. the NoP) can be a very significant limiting factor hindering the performance and efficiency of multi-chip AI accelerator architectures.

Given that multicast traffic is one of the sources of inefficiency in this context \cite{musavi2024, guirado2021dataflow}, wireless technology appears as a promising candidate to complement existing chiplet interconnects due to its low latency, reconfigurability, and inherent broadcast nature \cite{abadal2022graphene}. Antennas and transceivers operating at millimeter-wave frequencies can occupy less than 1 mm\textsuperscript{2}, operate at $\sim$1 pJ/bit and reach speeds in excess of 100 Gb/s \cite{yu2014architecture, tokgoz2018120gb, yi2021design}. In this context, building multi-chip AI accelerator architectures with wireless interconnects is expected to relieve the NoP bottleneck of existing accelerators in an effective and versatile way \cite{abadal2022graphene}. However, state-of-the-art developments either study how to improve a particular dataflow \cite{palesi2023wireless} or optimize the architecture template \cite{guirado2021dataflow} without making sure that (i) the mapping of the workloads on the architectures are optimal, and that (ii) the wireless link is used judiciously.

In this paper, we further delve into the study of cost-effective scalable wireless-enabled multi-chip AI accelerators with the aid of GEMINI~\cite{cai2024gemini}, which allows finding optimal mappings in multi-chiplet architectures. In particular, we modify the code of GEMINI to: (i) perform a brief study of the bottleneck of several AI workloads optimally mapped in multi-chiplet architectures, (ii) propose a reconfigurable wireless-enabled architecture capable of alleviating the NoP bottleneck across workloads, and (iii) assess the potential of our approach to increase the performance of multi-chiplet AI accelerators. We show an average (max) speedup of around 10\% (20\%) in a 3$\times$3 multi-chip architecture as well as the importance of balancing the load between the wired and wireless planes.

The remainder of the paper is organized as follows. In Section \ref{sec:related} we present the related work. Section \ref{sec:methodology} describes the proposed architecture to mitigate the workload-specific data movement bottleneck of existing architectures and outlines the methodology to evaluate it. In Section \ref{sec:results}, we discuss the results and in Section \ref{sec:conclusion} we conclude the paper.

\section{Related Work}
\label{sec:related}
\subsection{Design Space Exploration in Multi-Chip AI Accelerators}
Multi-chip AI accelerators have been proposed in multiple works. In SIMBA \cite{shao2019simba}, authors developed a multi-chip AI inference accelerator with a hierarchical interconnect architecture on a package. The objective was to enhance the energy efficiency and reduce the accelerator's inference latency by partitioning the non-uniform workload, considering communication-aware data placement, and implementing cross-layer pipelining.

In GEMINI \cite{cai2024gemini}, authors developed a scalable multi-chip AI inference accelerator design framework that explores the design space to deliver architectures for a particular workload and minimizes the monetary cost and the Energy-Delay Product (EDP). In more recent work, the authors in SCAR \cite{odema2024scar} developed a multi-chip multi-model AI inference accelerator that is scalable under heterogeneous traffic models (i.e., data center and AR/VR) with the objective of minimizing the EDP of the overall system.

Undoubtedly, these and similar multi-chip accelerator architectures have demonstrated excellent scalability and efficiency~\cite{das2024multi}. However, a significant gap remains in exploring the performance improvements achievable through wireless interconnects, which could address communication bottlenecks, improve energy efficiency, and enable more versatile designs.

\subsection{Wireless-enabled Multi-Chip AI Accelerators}
There are a few works that investigated wireless NoP to achieve performance enhancements in multi-chip AI accelerators. In \cite{palesi2023wireless}, the authors explored the use of wireless-enabled NoP in chiplet-based DNN accelerators to address the bottlenecks caused due to inter-chiplet communication. By leveraging single-hop communication and broadcast capabilities, wireless-enabled NoP significantly reduces latency and energy consumption, achieving superior EDP performance compared to traditional wired NoP. In WIENNA \cite{guirado2021dataflow}, the authors presented an NoP-based 2.5D DNN accelerator that addresses the bandwidth and scalability challenges of interposer-based designs. By utilizing wireless NoPs for high-bandwidth multicasting, WIENNA demonstrated significant improvements in both throughput and energy efficiency, providing a scalable solution for DNN acceleration. These works illustrated the potential of wireless-enabled AI accelerators but left several unexplored gaps in terms of optimal mapping or load balancing between wired and wireless interconnection networks.

In \cite{medina2023system}, authors explore the potential of in-package wireless communication as a scalable solution for multi-chiplet designs, reducing design complexity and freeing package resources by eliminating physical chiplet-to-substrate connections. Through simulations, it demonstrates that wireless interconnects can match or outperform wired alternatives, particularly for workloads like Convolutional Neural Networks (CNNs), with performance influenced by the wireless protocols and application mapping strategies. In this case, however, the AI workloads were mapped in general-purpose architectures.

\subsection{Workload Mapping on Multi-Chip AI Accelerators}
The mapper in SIMBA employed a layer-sequential mapping-first approach to reduce memory access overheads and data replication, with workload partitioning handled by Timeloop~\cite{parashar2019timeloop} for latency estimation and Accelergy~\cite{wu2019accelergy} for energy evaluation. The mapper employed in GEMINI~\cite{cai2024gemini} uses spatial-temporal mapping and inter-layer pipelining using SET~\cite{cai2023inter} mapper tool, and a cost model was customized to evaluate the cost of architecture. These enhancements significantly improved the overall performance of the system compared to the baseline mapping variants. The mapper used in SCAR is also based on SET~\cite{cai2023inter}, and the hybrid cost model is customized using MAESTRO~\cite{kwon2020maestro}. These works, particularly GEMINI, are leveraged in this paper to study the impact of wireless interconnects in optimally mapped AI workloads in multi-chip AI accelerators.

\section{Methodology}
\label{sec:methodology}
\begin{figure}[t]
\includegraphics[width=\columnwidth, trim=0cm 0.7cm 0cm 0cm]{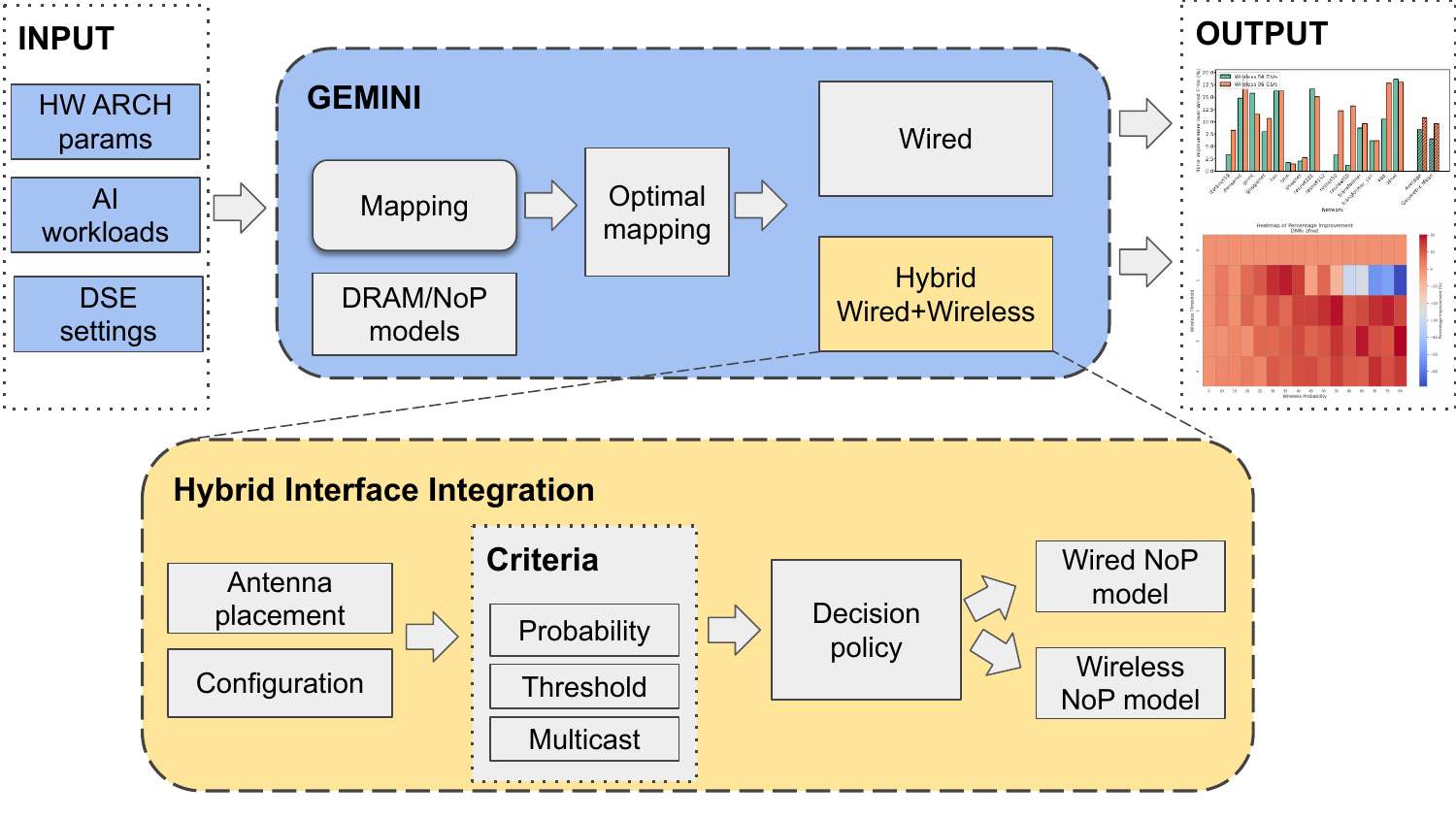}
\caption{Overall methodology. GEMINI is augmented with a wireless communication model and a wireless interface model. This allows to assess the impact of wireless interconnects on optimally mapped workloads in multi-chip AI accelerators.}
\label{fig:methodology}
\end{figure}

In this work, we extend the GEMINI framework \cite{cai2024gemini} by integrating wireless communication channels to alleviate the NoP load and improve the overall latency. This section outlines the modifications that have been made to the GEMINI simulator and the decision criteria to opt for the wireless channel, as illustrated in Figure~\ref{fig:methodology}. Also, Table~\ref{tab:simulationParameters-label} summarizes the values of the main simulation parameters.

\subsection{Architecture}
GEMINI is a mapping and architecture co-exploration framework for DNN inference chiplet accelerators. The original GEMINI architecture consists of multiple chiplets, each containing a mesh of PEs, and DRAM modules connected through wired communication links. Data transfer between chiplets and DRAM modules is done via the Die-to-Die (D2D) links, which can become a bottleneck for large-scale DNN workloads due to high communication overhead. On top of this architecture, illustrated in Figure \ref{fig:schematic}, we integrate wireless communication capabilities as described in the subsequent text. 

\subsection{Integration of Wireless Communication}
To address the communication bottleneck, we introduce a wireless communication channel into the GEMINI framework. Specifically, a wireless antenna and transceiver are placed on each chiplet as well as all DRAM modules. The antennas enable direct communication among chiplets and between chiplets and DRAMs, potentially reducing the number of hops required for message transfer and, consequently, the communication latency.

\subsubsection{Antenna Placement and Configuration}
The GEMINI simulator has been modified to include an antenna module representing wireless transceivers. We have strategically placed antennas at the center of each chiplet and DRAM module. The coordinates of the antennas are calculated based on the chiplet and DRAM positions to accurately model the physical layout. Therefore, the total number of antennas is equal to the sum of chiplets and DRAM modules. If a message needs to be wirelessly transmitted, it is routed through the NoC to the central routers of the chip, which are connected to the wireless interface.

\subsubsection{Wireless Communication Decision Criteria}
To decide whether to use wireless or wired communication for a given message, we implemented multiple configurable decision functions that consider the following factors:
\begin{itemize}
    \item \textbf{Multi-chip Multicast:} First, the network interface analyzes the destinations of a particular message. If the message is a multicast and there is at least one destination in a different chiplet than the source, wireless communication is used to exploit the broadcast nature of wireless channels.
    \item \textbf{Distance Threshold:} Second, a distance threshold is set to assess the minimum number of NoP hops required to consider wireless communication beneficial. If the number of chip-to-chip hops between the source and destination(s) exceeds this threshold, wireless communication is preferred.
    \item \textbf{Injection Probability:} Finally, a configurable probability is implemented to prevent a potential overuse of wireless channels. This probabilistic parameter ensures that the wireless channel is not saturated and became a source of bottleneck for workloads with a higher number of multi-chip and long-range multicasts.
\end{itemize}

\begin{table}[t]
  \centering
  \caption{Simulation Parameters}
  \begin{tabular}{l|l}
    \hline
    \textbf{Number of Chiplets} & \(3 \times 3\) \\
    \hline
    \textbf{DRAM Configuration} & 4 Chiplets, 16~GB/s per chiplet \\
    \hline
    \textbf{NoP Configuration} & XY Mesh, 32~Gb/s per side  \\
    \hline
    \textbf{NoC Configuration} & XY Mesh, 64~Gb/s per port \\
    \hline  \hline   
    \textbf{Wireless Bandwidth} & 64~Gb/s, 96~Gb/s \\
    \hline
    \textbf{Distance Threshold} & 1, 2, 3, 4 NoP hops \\
    \hline
    \textbf{Injection Probability} & 10\% to 80\% with step-size of 5\% \\
    \hline \hline
    \textbf{AI Workloads} & Darknet19, DenseNet, ZfNet, GNMT, Vgg,  \\
    & LSTM, ResNet50, ResNet101, ResNet152, \\
    & ResNeXt50, PNasNet, Transformer, \\
    & Transformer Cell, iRESNet, GoogleNet \\
    \hline 
  \end{tabular}
  \label{tab:simulationParameters-label}
\end{table}

\subsubsection{Wireless Communication Modeling}
When a message is designated for wireless transmission, it is loaded onto the source antenna and sent directly to the destination antennas, thereby inherently implementing broadcast or multicast functionality by virtue of wireless. The simulator updates the total wireless transmission and reception volumes. The impact on latency and energy consumption is computed based on these parameters.

\subsection{Simulator Modifications for Performance Evaluation}
GEMINI is not a cycle-accurate simulator rather employs certain approximations to speed up the simulation and allows fast mapping and exhaustive design space explorations. In particular, GEMINI calculates, layer by layer, computing time for each PE, the memory times for each DRAM chiplet, and interconnect times for each link i.e., NoC and NoP in an aggregated form. Then, it analyzes which element is the bottleneck for each layer. The total execution time is the sum of the maximum latency (i.e., that of the bottleneck) across all the layers of the workload. We also note that, GEMINI is not cycle-accurate, it does not take into consideration factors such as the contention in the NoP/NoC routers or in the DRAM chiplets.

To evaluate the benefits of wireless communication without altering the original simulation and mapping strategy of GEMINI, we simulate both wired and wireless communication paths for each message that qualifies for wireless transmission. We calculate the NoC and NoP hops for both cases and compute the difference to assess the performance improvements of having wireless, as described in subsequent section. 

\subsubsection{Wired Communication Path Simulation}
For messages that are sent via wireless channels, we also simulate the wired communication path to determine the number of hops and latency that would have occurred without wireless communication. This involves calling the standard unicast or multicast functions and accumulating the total hops and latency. This approach allows us to quantify the benefits of wireless communication by comparing against the baseline wired communication.

\subsubsection{Wireless Communication Path Simulation}
We simulate the wireless communication by updating the wireless-specific counters and latency calculations. This includes the total wireless transmission and reception volumes, as well as the wireless NoC hops. The simulator tracks the data sent and received via each antenna and models the wireless communication time by dividing the total volume of traffic by the wireless link bandwidth, similar to how GEMINI handles NoP/NoC times.

\section{Performance Evaluation}\label{sec:results}
\subsection{Experimental Setup}
We conducted experiments by varying key parameters affecting wireless communication to evaluate its impact on the overall performance. The simulation parameters are summarized in Table~\ref{tab:simulationParameters-label}. The additional parameters introduced for the wireless communication are the wireless bandwidth, with values that commensurate with state-of-the-art transceivers \cite{yu2014architecture,yi2021design}, the distance threshold, which is swept from 1 to 4 NoP hops, and the injection probability, which is swept from 10\% to 80\% with step-size of 5\% to assess the importance of load balancing in this scenario. 

We tested the modified GEMINI simulator using a set of representative DNN workloads that stress different aspects of the communication infrastructure. We choose benchmark models that contain multi-branch classic residual (e.g. ResNet50, ResNet152, GoogleNet, Transformer (TF), TF Cell) and inception (e.g. iRES) structures with more intricate dependencies which are prevalent and widely used in various scenarios such as image classification and language processing. These workloads differ in size and communication requirements, ensuring that the evaluation captures a wide range of realistic scenarios.

\begin{figure}[!t]
\includegraphics[width=\columnwidth]{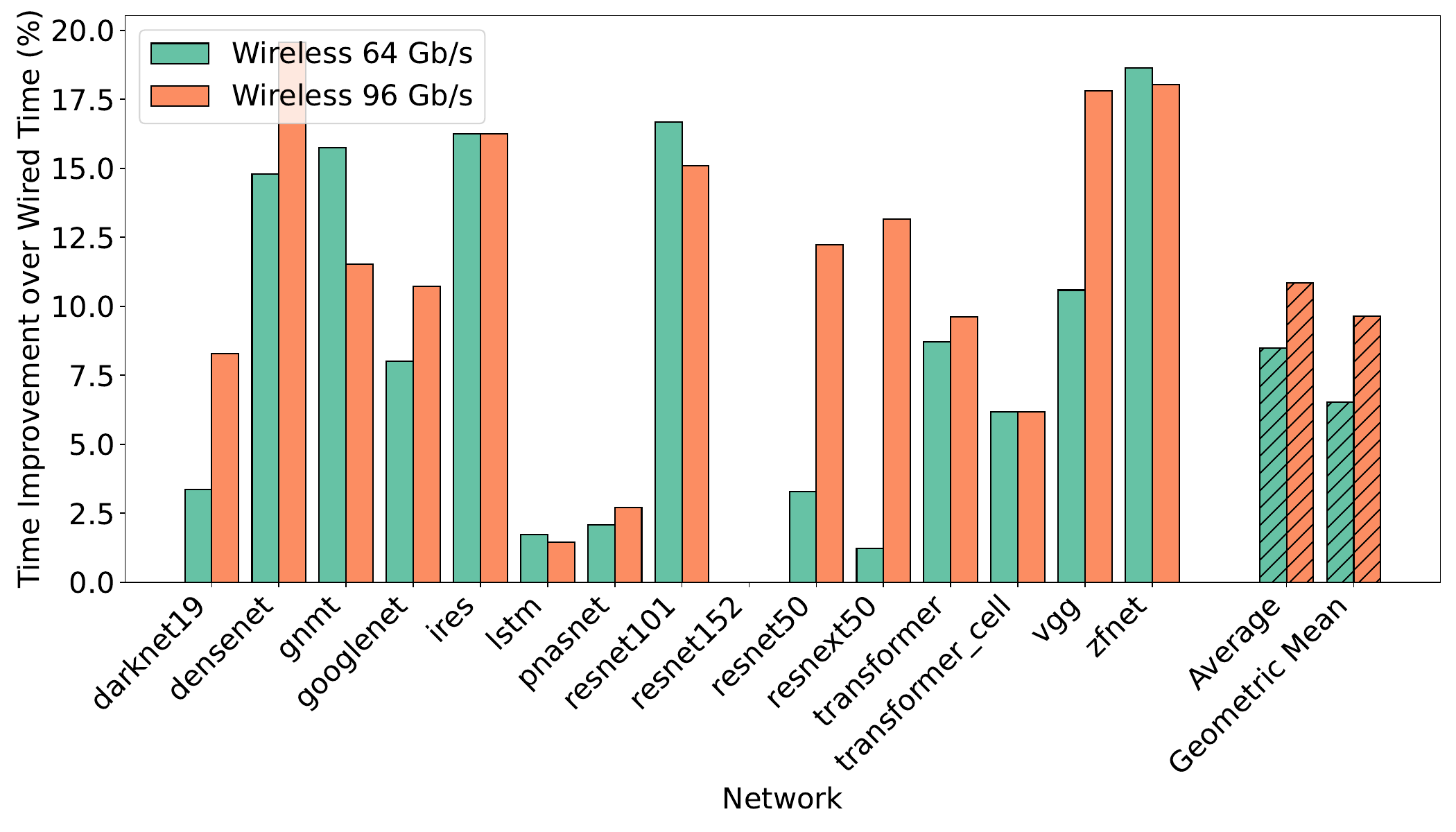}
\caption{Speedup of the proposed approach over a wired baseline in a 3$\times$3 multi-chip accelerator across the different evaluated AI workloads and for two different wireless bandwidths.}
\label{fig:improvement}
\end{figure}

\subsection{Results}
We compare the performance of the original GEMINI architecture with the wireless-enhanced version by analyzing the reduction in communication latency and energy consumption. The updated total hops and latency are computed by subtracting the wired communication metrics that were replaced by wireless communication. In our exploration, we sweep the distance threshold and injection probability parameters until finding a near-optimal value for each workload; this way, we assess the acceleration potential of the proposed approach. 

Figure~\ref{fig:improvement} shows the exploratory results for the two assumed wireless bandwidths. In particular, we plot the improvement of the hybrid wired-wireless architecture with respect to the wired baseline. There are several observations to be made from this figure:
\begin{itemize}
    \item The proposed approach improves performance across the board, except in the case of resnet152, which is mostly compute and NoC bound, as observed in Figure \ref{fig:bottleneck}. 
    \item The average speedups are around 7.5\% and 10\% for a wireless bandwidth of 64 Gb/s and 96 Gb/s, respectively, with maximum values of almost 20\%.
    \item In some cases, an increase in wireless bandwidth does not directly translate to an increase in the speedup. This could be due to the coarse exploration of the distance threshold and injection probability values, leading to a sub-optimal result for higher bandwidth values. This also suggests that the maximum attainable speedups might be higher than the ones shown here.
\end{itemize}

\begin{figure}[!t]
\includegraphics[width=\columnwidth, trim=0cm 0cm 0cm 0cm]{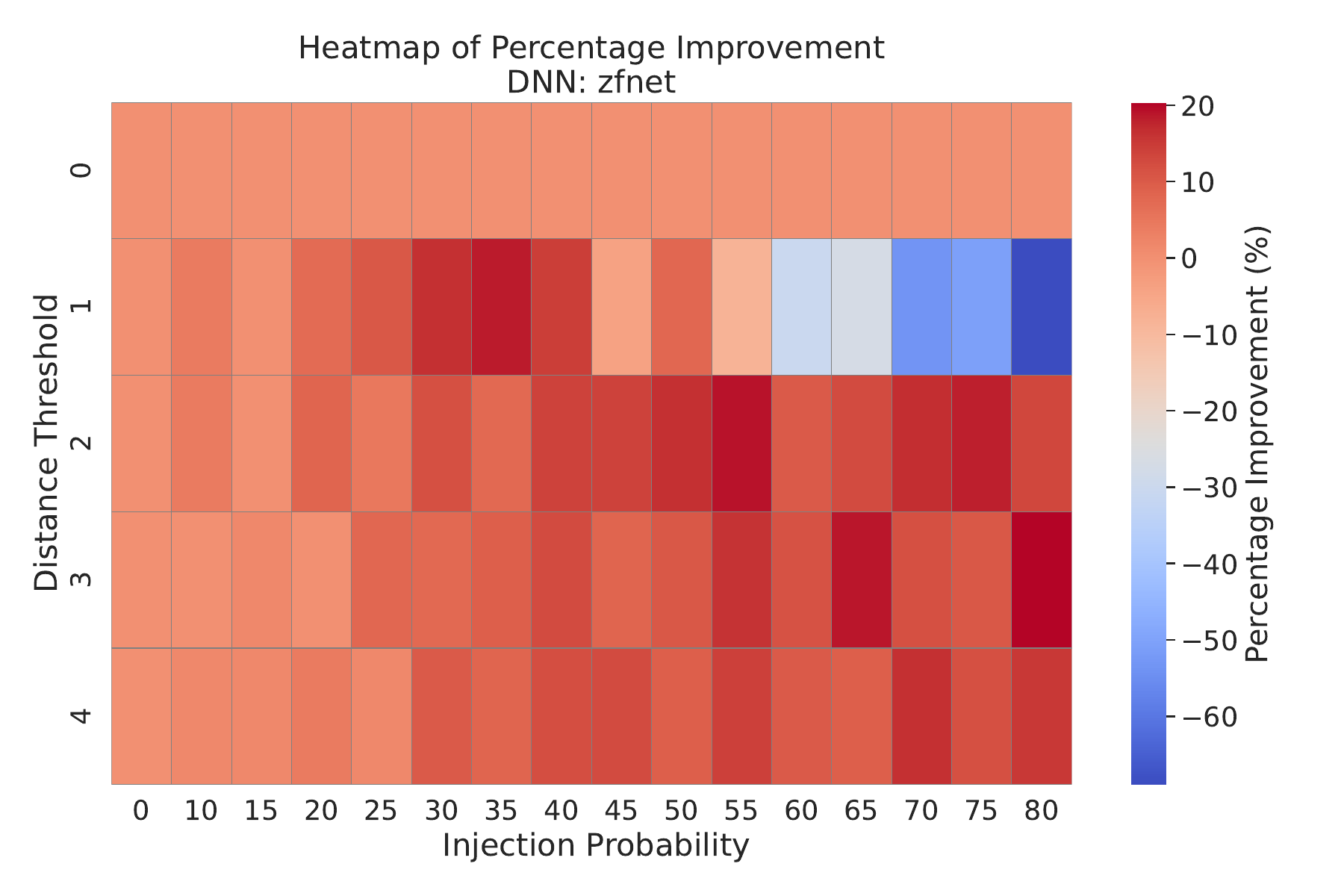}
\caption{Impact of the distance threshold and injection probability on the performance of the proposed approach for the \emph{zfnet} workload. Hotter colors indicate higher speedups whereas colder colors indicate performance degradations.}
\label{fig:exploration}
\end{figure}

To illustrate the importance of preventing the wireless network from becoming the bottleneck and the role of the distance threshold and injection probability in this regard, Figure~\ref{fig:exploration} shows the performance improvement (positive values) or degradation (negative values) in the \emph{zfnet} workload as a function of the wireless configuration parameters. Looking from left to right and top to bottom, the figure shows how increasing the load on the wireless link (by maintaining a low distance threshold and increasing the injection probability) can lead to a higher reward. However, the advantage is negated and turns into performance degradation for injection probabilities over 50\%. In that scenario, increasing the distance threshold can reduce the pressure on the wireless link and help regain the advantage of the wireless approach. This result, which changes from workload to workload, underscores the need for a mechanism to balance the load between the wired and wireless planes of the accelerator.

\section{Conclusion}
\label{sec:conclusion}
In this paper, we have showcased the potential of wireless interconnects for improving the performance and flexibility of multi-chip AI accelerators. We have first seen how the NoP can become a bottleneck in these scenarios. Based on our previous works, this is due to multicast patterns leading to congested bisection links, a situation that an overlaid wireless can help prevent. We have then seen how speedups of 20\% are achievable but contingent on finding a suitable load-balancing mechanism to prevent the wireless link from becoming saturated. In future work, we will investigate ways to proactively configure the wireless interface based on offline profiling of AI workloads and investigate alternative mapping methods capable to optimally exploit the advantages of the wireless interconnects.

\bibliographystyle{IEEEtran}
\bibliography{references}

\end{document}